\documentclass[12pt]{article}
\usepackage{graphicx}
\usepackage{graphicx}
\usepackage{latexsym}
\usepackage{amssymb}
\usepackage{amsmath}
\usepackage{color}
\newcommand{\eqn}[1]{&\hspace{-0.6em}#1\hspace{-0.6em}&}

\begin{document}
\baselineskip 0.6cm
%
\begin{titlepage}
\begin{center}

\begin{flushright}
SU-HET-13-2015
\end{flushright}

\vskip 2cm

{\Large \bf 
 A new dynamics of electroweak symmetry breaking with classically scale invariance
}

\vskip 1.2cm

{\large 
Naoyuki Haba$^1$, Hiroyuki Ishida$^1$, Noriaki Kitazawa$^2$, and \\
Yuya Yamaguchi$^{1,3}$
}

\vskip 0.4cm

$^1$Graduate School of Science and Engineering, Shimane University,\\
 Matsue 690-8504, Japan\\
$^2$Department of Physics, Tokyo Metropolitan University,\\
Hachioji, Tokyo 192-0397, Japan\\
$^3$Department of Physics, Faculty of Science, Hokkaido University,\\
 Sapporo 060-0810, Japan

\vskip 0.2cm


\vskip 2cm

\vskip .5in
\begin{abstract}
\noindent
We propose a new dynamics of the electroweak symmetry breaking  in a classically scale invariant version of the standard model. 
The scale invariance is broken by the condensations of additional fermions under a strong coupling dynamics.
The electroweak symmetry breaking is triggered by negative mass squared of the elementary Higgs doublet, 
which  is dynamically generated through the bosonic seesaw mechanism. 
We introduce a real pseudo-scalar singlet field interacting with additional fermions and Higgs doublet 
in order to avoid massless Nambu-Goldstone bosons from the chiral symmetry breaking in a strong coupling sector. 
We investigate the mass spectra and decay rates of these pseudo-Nambu-Goldstone bosons, 
and show they can decay fast enough without cosmological problems. 
We further evaluate the energy dependences of the couplings between elementary fields perturbatively, 
and find that our model is the first one which realizes the flatland scenario with the dimensional transmutation by the strong coupling dynamics. 
Similarly to the conventional flatland model with Coleman-Weinberg mechanism, the electroweak vacuum in our model is meta-stable. 
\end{abstract}
\end{center}
\end{titlepage}

\renewcommand{\thefootnote}{\#\arabic{footnote}} 
\setcounter{footnote}{0}

\section{Introduction}

The origin of the electroweak symmetry breaking (EWSB) remains a mystery. 
In the standard model (SM), 
the EWSB requires a negative mass squared for the Higgs doublet scalar field, whose magnitude is set by hand. 
We expect a fundamental theory which naturally gives the negative mass squared with the suitable value. 
In a model of supersymmetric extension of the SM, the EWSB can be realized by so-called radiative breaking~\cite{Inoue:1982pi}. 
However, the supersymmetry breaking scale must be high because of no signal of super-particle at any experiments so far. 
In technicolor (TC) model~\cite{technicolor}, the Higgs doublet field is no longer an elementary scalar field, 
and the EWSB is triggered by the techni-fermion condensation under strongly coupled TC gauge interaction. 
However, the naive TC model, which is just scale up of QCD, has already been excluded by the electroweak precision measurements.

Recently, there are a lot of studies of other possibilities to solve the gauge hierarchy problem 
by imposing a classically scale invariance with an additional $U(1)$ gauge symmetry~\cite{Hempfling:1996ht}-\cite{Haba:2015lka}. 
From the viewpoint of Bardeen's argument~\cite{Bardeen:1995kv}, 
we can only focus on logarithmic divergences, and the scale invariance protects large Higgs mass corrections. 
Under the classically scale invariance in terms of the cutoff regularization, 
the quadratic divergence itself can be subtracted by a boundary condition of the UV complete theory~\cite{Iso:2012jn}. 
Once we subtract the quadratic divergence from the theory, it never appears in the observables. 
In the model with an additional $U(1)$ gauge symmetry, 
the scale invariance is broken by the Coleman-Weinberg mechanism~\cite{Coleman:1973jx}, 
and if the breaking scale is not so far from the electroweak (EW) scale, there is no gauge hierarchy problem. 
On the other hand, a strong coupling dynamics can also realize such an EWSB with classically scale invariance~\cite{Hur:2007uz,Holthausen:2013ota}, 
where an additional singlet scalar mediates dimensional transmutation in the strong coupling sector to the SM sector. 
However, the sign of the coupling between the Higgs doublet and the additional scalar is assumed to be negative, 
so that the negative mass squared of the Higgs doublet is realized. 
Therefore, the origin of the EWSB is not necessary and inevitable in this scenario, 
and we are going to try the dynamical realization of negative mass squared by the bosonic seesaw mechanism~\cite{BSS}.

In this paper,  we expand the SM gauge group by $SU(N_{\rm TC})$ technicolor gauge symmetry with the classically scale invariant framework. 
The techni-fermions, which belong to vector-like representations under TC gauge symmetry as well as electroweak gauge symmetry, 
are introduced. 
Though the chiral symmetry breaking happens by techni-fermion condensations, 
the EWSB does not happen by this strong coupling TC dynamics itself. 
We show that the EWSB dynamically occurs in an inevitable way by the bosonic seesaw mechanism 
between the elementary Higgs scalar field and a composite scalar field.
To avoid massless Nambu-Goldstone (NG) bosons by the chiral symmetry breaking in strong coupling sector, 
we introduce a real pseudo-scalar singlet field and its interactions with techni-fermions and Higgs doublet. 
We analyze the mass spectrum of the pseudo-NG (pNG) bosons and estimate their decay rates. 
We show that the pNG bosons can decay fast enough to avoid cosmological problems. 
We further show that our model can be regarded as the first model of the flatland scenario with strong coupling dynamics.
All three coupling constants in scalar potential can vanish at the Planck scale.
Similarly to the conventional flatland model with Coleman-Weinberg mechanism, the EW vacuum in our model is meta-stable.

\section{Bosonic seesaw mechanism}

By imposing classically scale invariance, the mass term of the Higgs potential is forbidden and the Higgs potential becomes 
\begin{eqnarray}
V \eqn{=} \lambda 
\left( 
H^\dagger H
\right)^2\,. \label{Eq:pot_ori}
\end{eqnarray}
The EWSB does not occur by this potential, and we try to use the dimensional transmutation in the strong coupling sector, 
where there are two vector-like techni-fermions as shown in Tab.~\ref{Tab:Model}. 
\begin{table}
\centering
\begin{tabular}{|c||c|c|c|}
\hline
 &$SU (N_{\rm TC})$ &$SU(2)_L$ &$U(1)_Y$\\
\hline \hline
$H$ &$1$ &$2$ &$1/2$\\
\hline 
$\chi$ &$N_{\rm TC}$ &$2$ &$-1/2$\\
\hline
$\psi$ &$N_{\rm TC}$ &$1$ &$0$\\
\hline
\end{tabular}
\caption{Charge assignments of techni-fermions and the Higgs doublet.}\label{Tab:Model}
\end{table}
Due to the classically scale invariance, vector-like fermion masses are also forbidden. 
In the model, the chiral symmetry in the strong coupling sector 
$SU(3)_{L} \times SU(3)_{R} \times U(1)_A$ is explicitly broken by the SM gauge symmetry, $SU(2)_L \times U(1)_Y$, 
 and the remaining symmetry is $SU(2)_{\chi_L} \times SU(2)_{\chi_R} \times U(1)_{\chi_A} \times U(1)_{\psi_A}$.
There is also $U(1)_{\chi_V}\times U(1)_{\psi_V}$, 
which is similar to the baryon number symmetry.\footnote{
They guarantee the stability of the lightest techni-baryon 
 which can be a candidate of the dark matter.
(For instance, see Ref.~\cite{Antipin:2015xia}.)}
This vector-like symmetry is expected to be unbroken by the strong-coupling technicolor dynamics 
 due to the Vafa-Witten's theorem~\cite{Vafa:1984xg}. 
The chiral symmetry should be broken as preserving $SU(2)_L \times U(1)_Y$ symmetry, 
then we expect $\langle \bar{\chi} \psi \rangle = \langle \bar{\psi} \chi \rangle = 0$ and $\langle \bar{\chi} \chi \rangle \neq 0$, 
 $\langle \bar{\psi} \psi \rangle \neq 0$. 
They cause chiral symmetry breaking 
$SU(2)_{\chi_L} \times SU(2)_{\chi_R} \times U(1)_{\chi_A} \times U(1)_{\psi_A} \to SU(2)_{\chi_V}$.
There are five NG bosons; two massive pNG bosons of anomalous $U(1)_{\chi_A}$ and $U(1)_{\psi_A}$ breakings, 
and three massless NG bosons corresponding to the breaking $(SU(2)_{\chi_L} \times SU(2)_{\chi_R})/SU(2)_{\chi_V}$ symmetry. 
If we neglect $SU(2)_L \times U(1)_Y$, the chiral symmetry breaking of $SU(3)_L \times SU(3)_R \times U(1)_A \to SU(3)_V$ occurs.
There are nine NG bosons; one massive pNG boson of $U(1)_A$ breaking, and eight massless NG bosons.

The techni-fermions interact with Higgs doublet $H$ through the Yukawa interactions, 
\begin{eqnarray}
-\mathcal{L}_{\rm Yukawa} \eqn{=} 
 y_L \bar{\chi}_L H \psi_R + y_R \bar{\chi}_R H \psi_L + {\rm h.c.}\,.
\label{Eq:Yukawa}
\end{eqnarray}
After the techni-fermion condensation, $\chi_{L\,,R}$ and $\psi_{L\,,R}$ are confined by non-perturbative effects, 
and $\bar{\chi}_L \psi_R$ and $\bar{\chi}_R \psi_L$ couple to a ``meson'' state, that is just a composite Higgs doublet,
 $\Theta \sim \bar{\chi} \psi/\Lambda_{\rm TC}^2$. 
When $y_L$ and $y_R$ are real, there is the charge conjugation invariance. 
Here, we assume $y_L = y_R =y$ for simplicity. 
The Yukawa interactions in Eq.(\ref{Eq:Yukawa}) are CP invariant in this case. 
The composite Higgs doublet mixes with the elementary Higgs doublet, 
and the mass matrix becomes 
\begin{eqnarray}
-\mathcal{L}_{\rm mass}
\eqn{=}
\begin{pmatrix}
H^\dagger \ \ \Theta^\dagger
\end{pmatrix}
\begin{pmatrix}
0 &y \Lambda_{\rm TC}^2\\
y \Lambda_{\rm TC}^2 &\alpha \Lambda_{\rm TC}^2\\
\end{pmatrix}
\begin{pmatrix}
H \\
\Theta
\end{pmatrix} \label{Eq:Higgs_mass}\\
\eqn{\simeq}
\begin{pmatrix}
H_1^\dagger \ \ H_2^\dagger
\end{pmatrix}
\begin{pmatrix}
-\frac{y^2}{\alpha}\Lambda_{\rm TC}^2 & 0 \\
0 &\alpha \Lambda_{\rm TC}^2\\
\end{pmatrix}
\begin{pmatrix}
H_1 \\
H_2
\end{pmatrix}\,,
\end{eqnarray}
where $\alpha$ is a dimensionless positive coefficient of $\mathcal{O}(1)$. 
Here, $y \ll \alpha$ is assumed, since the chiral symmetry breaking terms should be small to be treated perturbatively. 
As we will see later, the small Yukawa coupling $y$ is also necessary for the hierarchy between the EW and TC condensation scales. 
As a result, the lighter (heavier) mass eigenstate $H_1$ ($H_2$) is almost $H$ ($\Theta$). 
The field $H_1$ is regarded as the SM-like Higgs doublet, 
and the negative mass squared is dynamically obtained through the bosonic seesaw mechanism. 
The field $H_2$ has mass of ${\cal O}(\Lambda_{\rm TC})$.


There are massless NG bosons in the present stage. 
To avoid the massless NG bosons, we introduce real pseudo-scalar field, $S$, which has interactions, 
\begin{eqnarray}
-\mathcal{L}_S \eqn{=} 
 g_S S \bar{\chi} i \gamma_5 \chi + g'_S S \bar{\psi} i \gamma_5 \psi\,, \label{Lag:pseudo}
\end{eqnarray}
where $g_S$ and $g'_S$ are taken to be real to keep the CP invariance. 
Since we can expect 
$\langle \bar{\chi} i \gamma_5 \chi \rangle =0$ and $\langle \bar{\psi} i \gamma_5 \psi \rangle=0$ in vector-like technicolor dynamics, 
the no tadpole term of $S$ is not generated. 

Now the potential in Eq.(\ref{Eq:pot_ori})
 is modified as 
\begin{eqnarray}
V_{\rm eff} \eqn{=} 
\lambda \left( H^\dagger H \right)^2 
+
\kappa S^2 H^\dagger H 
+ 
\lambda_S S^4 
+ 
y \Lambda_{\rm TC}^2 \left( H^\dagger \Theta + \Theta^\dagger H \right) 
+ 
\alpha \Lambda_{\rm TC}^2 \Theta^\dagger \Theta\,,
\end{eqnarray}
where fourth and fifth terms are obtained from Eq.~(\ref{Eq:Higgs_mass}). 
Since we have assumed the hierarchy between the light and heavy mass eigenstates, 
the heavier mass eigenstate $H_2$ is decoupled at low energies. 
Therefore, the effective potential at low energy is 
\begin{eqnarray}
V_{\rm eff} \eqn{\simeq} 
\lambda \left( H_1^\dagger H_1 \right)^2 
+
\kappa S^2 H_1^\dagger H_1 
+ 
\lambda_S S^4 
- 
\frac{y^2}{\alpha} \Lambda_{\rm TC}^2 H_1^\dagger H_1 
- 
\frac{1}{2} m_S^2 S^2 
\,, 
\label{Eq:Veff}
\end{eqnarray}
where we include the mass term of $S$ which is generated by bosonic seesaw mechanism again. 
We will give an analysis about this issue shortly in the next section.

The vacuum expectation values (VEVs) of $H_1$ and $S$ can be evaluated by the effective potential Eq.~(\ref{Eq:Veff}). 
The stationary conditions are 
\begin{eqnarray}
\left( \lambda v_H^2 + \kappa v_S^2 - \frac{y^2}{\alpha} \Lambda_{\rm TC}^2 \right) v_H \eqn{=} 0 \,,\\
\left( \kappa v_H^2 + 4 \lambda_S v_S^2 - m_S^2 \right) v_S \eqn{=} 0\,,
\end{eqnarray}
where $\langle H_1 \rangle = (0, v_H/\sqrt{2})^{\rm T}$ and $\langle S \rangle = v_S$. 
Note that $v_H$ should corresponds to the EW scale ($v_H=246\,{\rm GeV}$), and nonzero $v_S$ causes spontaneous CP violation. 
Except for a trivial solution $v_H = v_S = 0$, there are three possibilities of solutions as follows. 
\begin{itemize}
\item $v_H =0$ and $v_S \neq 0$

In this case the EW symmetry is not broken, while $v_S$ can be estimated as 
\begin{eqnarray}
v_S^2 
\eqn{=} 
\frac{m_S^2}{4 \lambda_S}\,,
\end{eqnarray}
where $m_S^2$ must be positive. 
To satisfy $v_H=0$, i.e., to realize the positive mass squared of $H_1$, the following condition must be satisfied:
\begin{eqnarray}
\kappa v_S^2 - \frac{y^2}{\alpha} \Lambda_{\rm TC}^2 > 0 \,.
\end{eqnarray}
Thus, a certain large value of $\kappa$ is required when $v_S$ is $\mathcal{O} (\Lambda_{\rm TC})$. 
Anyway, we do not consider this case, since the EW symmetry is unbroken.

\item $v_H \neq 0$ and $v_S =0$

In this case the EWSB occurs and its scale is given by
\begin{eqnarray}
v_H^2 \eqn{=} 
\frac{y^2}{\lambda \alpha} \Lambda_{\rm TC}^2\,.
\end{eqnarray}
This is really a solution if the mass squared of $S$ is positive, that is,
\begin{eqnarray} 
\frac{\kappa y^2}{\lambda \alpha} \Lambda_{\rm TC}^2 - m_S^2 > 0\,.
\end{eqnarray}
This condition is always satisfied for a sufficiently large $\kappa$. 
Since we would like to treat $\kappa$ perturbatively in good approximation, we do not adopt this case also. 

\item $v_H \neq 0$ and $v_S \neq 0$ 

This case leads a suitable result.
The stationary conditions give 
\begin{eqnarray}
v_H^2 \eqn{=} 
\frac{1}{4 \lambda \lambda_S - \kappa^2} 
\left( 
\kappa m_S^2 + 4 \lambda_S \frac{y^2}{\alpha}  \Lambda_{\rm TC}^2 
\right)\,,\\
v_S^2 \eqn{=} 
\frac{1}{4 \lambda \lambda_S - \kappa^2} \left( \lambda m_S^2 - \kappa \frac{y^2}{\alpha} \Lambda_{\rm TC}^2 \right)\,.
\end{eqnarray}
Since the squared VEVs must be positive, a certain small value of $\kappa$ are required. 
In the limit of $\kappa \to 0$, the VEVs are approximately given by 
\begin{eqnarray}
v_H^2 \eqn{\simeq} 
\frac{y^2}{\lambda \alpha} \Lambda_{\rm TC}^2\,,\qquad
v_S^2 \simeq 
\frac{1}{4 \lambda_S} m_S^2\,,\label{Eq:VEV_EW}
\end{eqnarray}
where $m_S^2$ must be positive. 
Since $S$ obtains a nonzero VEV, a mixing term with Higgs doublet affects the Higgs mass through $\kappa |H_1|^2 S^2$. 
However, it is negligible because $\kappa$ is assumed to be sufficiently small. 
(In the case of $\kappa \simeq 0$, we can treat $H_1$ and $S$ independently.) 
From now on, we adopt this case with taking sufficiently small value of $\kappa$. 
\end{itemize}

\section{Mass spectra and decay rates of pNG bosons}

Now let us investigate the mass spectra and decay rates of the pNG bosons.
Actually, all nine NG bosons should become massive due to the introduction of $S$, since the chiral symmetries
 $SU(2)_{\chi_L} \times SU(2)_{\chi_R} \times U(1)_{\chi_A} \times U(1)_{\psi_A}$ is explicitly broken down into $SU(2)_{\chi_V}$ 
by the interactions of Eq.~(\ref{Lag:pseudo}).
The results of the mass spectra are summarized in Tab. \ref{Tab:TP}.  
\begin{table}
\centering
\begin{tabular}{|c||c|c|c|c|}
\hline
 &Operators &$SU(2)_L$ &$U(1)_Y$ & Masses \\
\hline
$\eta_\chi$ &$\bar{\chi} i \gamma_5 \chi$ &$1$ &$0$ & $\beta \Lambda_{\rm TC}^2$\\
\hline
\rule{0cm}{0.45cm}$\eta_\psi$ &$\bar{\psi} i \gamma_5 \psi$ &$1$ &$0$ & $\beta \Lambda_{\rm TC}^2$\\
\hline
$\Pi_i$ ($i=1, 2, 3$) &$\bar{\chi} i \gamma_5 \sigma_i \chi$ &$3$ &$0$ & $\frac{8 \pi^2 g_S^4}{\lambda_S \beta^2} \Lambda_{\rm TC}^2$ \\
\hline
\rule{0cm}{0.75cm}$\Sigma = \genfrac{(}{)}{0pt}{0}{\Sigma^0}{\Sigma^-}$ &$\bar{\psi} i \gamma_5 \chi$ &$2$ &$-1/2$ & $\frac{16 \pi^2 g_S^4}{\lambda_S \beta^2} \Lambda_{\rm TC}^2$ \\
\hline
\rule{0cm}{0.75cm}$\bar{\Sigma} = \genfrac{(}{)}{0pt}{0}{\bar{\Sigma}^0}{\bar{\Sigma}^+}$ &$\bar{\chi} i \gamma_5 \psi$ &$2$ &$1/2$ & $\frac{16 \pi^2 g_S^4}{\lambda_S \beta^2} \Lambda_{\rm TC}^2$ \\
\hline
\end{tabular}
\caption{Summary of the pNG bosons. $\sigma_i$ are Pauli matrices.}\label{Tab:TP}
\end{table}

\vspace{3mm}

First, we investigate pNG boson mass spectra. 
The SM singlet pNG bosons ($\eta_\chi$ and $\eta_\psi$) mix with $S$, and the mass matrix is written by 
\begin{eqnarray}
-\mathcal{L}_{S \mathchar`- \eta_\chi \mathchar`- \eta_\psi} \eqn{=} 
\frac{1}{2} 
\begin{pmatrix}
S \ \ \eta_\chi^\dagger \ \ \eta_\psi^\dagger
\end{pmatrix} 
\begin{pmatrix}
0 &g_S \Lambda_{\rm TC}^2 &g'_S \Lambda_{\rm TC}^2\\
g_S \Lambda_{\rm TC}^2 &\beta_\chi \Lambda_{\rm TC}^2 &0\\
g'_S \Lambda_{\rm TC}^2 &0 & \beta_\psi \Lambda_{\rm TC}^2
\end{pmatrix}
\begin{pmatrix}
S \\
\eta_\chi \\
\eta_\psi
\end{pmatrix}\,, \label{Eq:light-mix}
\end{eqnarray}
where $\beta_\chi$ and $\beta_\psi$ are dimensionless positive coefficients of  $\mathcal{O}(1)$. 
All off-diagonal elements are induced from Eq.~(\ref{Lag:pseudo}). 
The determinant of this mass matrix is
 $- (g_S^2 \beta_\psi + g_S^{'2} \beta_\chi)\Lambda_{\rm TC}^6 <0$, 
 thus $S$ has a negative mass term. 
Taking $g_S = g'_S \ll \beta_\chi = \beta_\psi = \beta$,
 for simplicity,  
 mass eigenvalues of $S$, $\eta_\chi$, and $\eta_\psi$
 can be estimated as 
\begin{eqnarray}
- m_S^2 \eqn{\simeq} 
- \frac{2 g_S^2}{\beta} \Lambda_{\rm TC}^2\,,\qquad
m_{\eta_\chi}^2 = m_{\eta_\psi}^2 \simeq 
\beta \Lambda_{\rm TC}^2\,,
\label{ps_mass}
\end{eqnarray}
respectively.
The smallness of $g_S$ is natural, since it is expected to break the chiral symmetry perturbatively. 
Note that $S$ has the negative mass term by the bosonic seesaw mechanism again (See Eq.~(\ref{Eq:Veff}). 

Using Dashen's formula~\cite{Dashen:1969eg} and VEVs of $H_1$ and $S$ in Eq.(\ref{Eq:VEV_EW}), 
the masses of $\Pi$ and $\Sigma$ are estimated as 
\begin{eqnarray}
m_{\Pi}^2 f_\Pi^2 
\eqn{=} 
\langle 0 | \left[ Q,\, \left[ Q,\, \mathcal{H}_S \right] \right] | 0 \rangle 
\simeq
\frac{8 \pi^2 g_S^4}{\lambda_S \beta^2} \Lambda_{\rm TC}^4\,,\\
m_{\Sigma}^2 f_\Sigma^2 
\eqn{=} 
\langle 0 | \left[ Q,\, \left[ Q,\, \mathcal{H}_S \right] \right] | 0 \rangle
\simeq 
\frac{g_S^4}{\lambda_S \beta^2} \Lambda_{\rm TC}^4\,,
\end{eqnarray}
where $\mathcal{H}_S = g_S S \bar{\chi} i \gamma_5 \chi + g_S S \bar{\psi} i \gamma_5 \psi$ from Eq.~(\ref{Lag:pseudo}) 
and, $f_\Pi$ and $f_\Sigma$ are decay constants of $\Pi$ and $\Sigma$, respectively. 
Both decay constants are evaluated by naive dimensional analysis~\cite{Manohar:1983md,Georgi:1986kr} 
as $\Lambda_{\rm TC} \simeq 4 \pi f_{\Pi,\,\Sigma}$ by analogy with QCD.
Therefore, the masses of $\Pi$ and $\Sigma$ are estimated as 
\begin{eqnarray}
m_\Pi^2 
\eqn{\simeq} 
\frac{8 \pi^2 g_S^4}{\lambda_S \beta^2} \Lambda_{\rm TC}^2\,,\qquad
m_\Sigma^2 \simeq
\frac{16 \pi^2 g_S^4}{\lambda_S \beta^2} \Lambda_{\rm TC}^2\,.
\end{eqnarray}
In the following, we take $\Lambda_{\rm TC} = 10\,{\rm TeV}$ and $\alpha = \beta =1$ for an explicit example. 
Then, the coupling $y$ is evaluated as $y \simeq 0.068$ from Eq.~(\ref{Eq:VEV_EW}).
If we also take $\lambda_S=10^{-3}$ and $g_S = 0.05$, 
$v_S $ and the mass of $\Pi$ and $\Sigma$ are evaluated 
as $v_S \simeq 11\,{\rm TeV}$, $m_\Pi \simeq 7\,{\rm TeV}$ and $m_\Sigma \simeq 10\,{\rm TeV}$, respectively.

\vspace{3mm}

Next, we estimate decay rates of the pNG bosons by analogy with light mesons in QCD. 
A charged components of $\Sigma$ and $\Pi$ can decay into their neutral components 
and the SM fermions through the weak interactions. 
$\eta_\chi$ and the neutral component of $\Pi$ ($\Pi^0$) decay into two photons by analogy with $\pi^0$ decay in the SM. 
The decay rate of $\eta_\chi$ is evaluated as 
\begin{eqnarray}
\Gamma ( \eta_\chi \to \gamma \gamma ) 
\eqn{=} 
\left( \frac{N_{\rm TC} e^2}{4 \pi^2 f_{\eta_\chi}} \right)^2 \frac{m_{\eta_\chi}^3}{64 \pi}
\simeq
\frac{N_{\rm TC}^2 \alpha_{\rm em}^2}{4 \pi} \frac{m_{\eta_\chi}^3}{\Lambda_{\rm TC}^2}\,,
\end{eqnarray}
where we have used $f_{\eta_\chi} \simeq \Lambda_{\rm TC} /4 \pi$
 and $\alpha_{\rm em}=e^2/4\pi$. 
When we take $N_{\rm TC} =3$ for example, 
the decay rate is estimated by $\Gamma ( \eta_\chi \to \gamma \gamma ) \simeq 400\,{\rm MeV}$. 

The neutral component of $\Sigma$ ($\Sigma^0$) also decays into two photons via a mixing with $\eta_\chi$. 
The effective $\Sigma$-$\eta_\chi$ mixing is evaluated by Dashen's formula as 
\begin{eqnarray}
m_{\Sigma \mathchar`- \eta_\chi}^2 
\eqn{\simeq} 
(4 \pi)^2 \sqrt{2} y v_H \Lambda_{\rm TC}\,,
\end{eqnarray}
and hence, the magnitude of $\Sigma$-$\eta_\chi$ mixing is given by 
\begin{eqnarray}
V_{\Sigma \mathchar`- \eta_\chi} \equiv
\frac{m_{\Sigma \mathchar`- \eta_\chi}^2}{m_\Sigma^2} 
\eqn{\simeq} 
\frac{\sqrt{2} y \lambda_S \beta^2}{g_S^4} \frac{v_H}{\Lambda_{\rm TC}}\,. 
\end{eqnarray}
Thus, we obtain $V_{\Sigma \mathchar`- \eta_\chi} \simeq 0.4$
 by using the same numerical values as above. 
As a result, we find $\Gamma (\Sigma^0 \to \gamma \gamma) \simeq V_{\Sigma \mathchar`- \eta_\chi}^2 \times \Gamma ( \eta_\chi \to \gamma \gamma ) \simeq 60\,{\rm MeV}$.

The decay mode of the lightest neutral pNG boson $\eta_\psi$ is a little bit tricky. 
The decay process is $\eta_\psi \to S \to \eta_\chi \to \gamma \gamma$ through mass mixings.
Therefore, the lifetime of $\eta_\psi$ would be the longest among the pNG bosons. 
Since $S$-$\eta_\psi$ and $S$-$\eta_\chi$ effective mixing couplings can be evaluated from Eq.~(\ref{Eq:light-mix}) as $g_S'/\beta_\psi$ and $g_S/\beta_\chi$, respectively, the decay rate can be estimated as 
\begin{eqnarray} 
\Gamma ( \eta_\psi \to \gamma \gamma ) \eqn{\simeq} 
\left( \frac{g_S'}{\beta_\psi} \frac{g_S}{\beta_\chi} \right)^2 \times \Gamma ( \eta_\chi \to \gamma \gamma )\,. 
\end{eqnarray}
Thus, $\Gamma ( \eta_\psi \to \gamma \gamma )$ is around $3\,{\rm keV}$ using the same numerical values as above.
Even the lightest pNG boson can decay much faster than the QCD neutral pion
 ($\Gamma ( \pi^0  \to \gamma \gamma) \simeq 7.7\,{\rm eV}$). 
As a result, we can expect that all the pNG bosons decay into the SM particles fast enough without cosmological problems.

\section{Viewpoint from flatland scenario}

Our model is classically scale invariant. 
Now, as an interesting possibility, let us consider whether the scalar potential vanishes at the Planck scale. 
This constraint is severer than the scale invariant condition,
and all the scalar quartic couplings must be zero at the Planck scale.  
This situation has been studied in so-called flatland scenario~\cite{Iso:2012jn,Chun:2013soa,Hashimoto:2013hta,Hashimoto:2014ela},
where a singlet scalar field has an interaction with the SM Higgs doublet, 
and its VEV induces negative mass squared of the Higgs~\cite{Iso:2009ss,Iso:2012jn,Chun:2013soa,Hashimoto:2013hta,Hashimoto:2014ela,Guo:2015lxa}.  
The mass scale is determined by the singlet VEV, which is generated by Coleman-Weinberg mechanism. 
This scenario can induce not only the VEV of the scalar field but also the negative mass squared of the Higgs doublet. 
On the other hand, another dimensional transmutation mechanism by using strong coupling  dynamics has been proposed  in~\cite{Hur:2007uz,Holthausen:2013ota}.  
These models do not follow the flatland scenario, since scalar couplings do not vanish at the Planck scale. 
In these models the EWSB is not necessary and inevitable, 
because we need to choose the sign of the coupling appropriately.
It is worth noting that our model can inevitably induce the negative mass squared of the Higgs doublet by the strong coupling dynamics, 
and also all the scalar couplings can vanish at the Planck scale simultaneously.
These are remarkable points. 

\begin{figure}
\begin{center}
\includegraphics[width=8cm,clip]{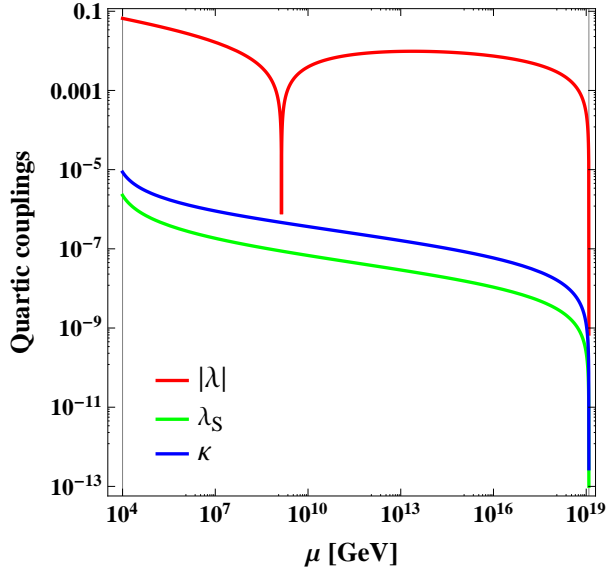}
\caption{The running of quartic couplings between $\Lambda_{\rm TC}$ and the Planck scale which are denoted by the black lines.
We take $\Lambda_{\rm TC} = 10 {\rm TeV}$.
All quartic couplings are zero at the Planck scale. 
}
\label{Fig_quartics}
\end{center}
\end{figure}

The running of the scalar quartic couplings in our model are shown in Fig.~\ref{Fig_quartics}. 
Here we have taken the top quark mass as $174.7 {\rm GeV}$ as a reference value, 
which is slightly heavier than the central value from the collider experiment~\cite{ATLAS:2014wva}. 
The curve of $|\lambda|$ is concave down between $10^9$ GeV and just below the Planck scale, where $\lambda <0$ in actually.  
This means that there is a lower-energy vacuum than the EW one. 
Thus, imposing the vanishing potential at the Planck scale makes the Higgs quartic coupling negative below the Planck scale, 
and the EW vacuum is meta-stable.  
This situation is the same as the flatland model~\cite{Haba:2015rha}.   
It is because the magnitude and energy scale dependence of $\lambda$ around the Planck scale are almost same as the SM. 
Although there seem to be some degrees of freedom in parameters of the bosonic seesaw mechanism, 
there is no room of adjustment, 
since the value of negative mass squared at the EW scale must be fixed to reproduce the SM. 
Thus our model cannot escape from the meta-stable vacuum, 
but the EW vacuum is stable enough compared with the age of our universe~\cite{Rose:2015fua}.\footnote{
Our analysis has been done at one-loop level, but the result does not change significantly from the analysis including higher orders.}  
Our model is the first one which realizes the flatland scenario through the dimensional transmutation of the strong coupling dynamics, in which we can evaluate the energy dependence of the scalar couplings perturbatively. 
Note that the larger $y$ could make the vanishing potential realized with the central value of the top quark mass. 
It leads smaller $\Lambda_{\rm TC}$ and lighter the pNG bosons.

\section{Discussions and conclusions}

The origin of the EWSB is not established yet, although the SM-like Higgs boson has been discovered. 
In this paper, we have investigated the dynamical origin of the EWSB via the bosonic seesaw mechanism 
in a classically scale invariant version of the SM. 
We have introduced the $SU(N_{\rm TC})$ technicolor gauge symmetry for the dimensional transmutation 
by the techni-fermion condensations. 
In this model, the mixing between the elementary and composite Higgs doublets becomes the origin of EWSB.
An extra real pseudo-scalar singlet field has also been introduced to avoid massless NG bosons. 
We have estimated mass spectra and decay rates of the pNG bosons. 
We have checked that all of the pNG bosons can decay fast enough without cosmological problems. 
Our model is the first model which realizes the flatland scenario through the dimensional transmutation of the strong coupling dynamics, 
in which we can evaluate the energy dependence of the scalar couplings perturbatively. 
Similarly to the conventional flatland model with Coleman-Weinberg mechanism, 
the EW vacuum in our model is meta-stable. 

Finally, we comment on the collider phenomenology. 
When the singlet pseudo-scalar is light enough to be produced at the collider, 
some vestiges could be searched in the future collider experiments. 
In addition, since there can exist light new mesons depending on the parameters, 
they might be detectable at collider experiments.

\subsection*{Acknowledgement}

This work is partially supported by Scientific Grant by the Ministry of Education, Culture,
Sports, Science and Technology, Nos. 24540272, 26400253, and 26247038. The works of N.H. and Y.Y. are
supported by Scientific Grant by the Ministry of Education, Culture, Sports, Science
and Technology (No. 15H01037) and Research Fellowships of the Japan Society for the Promotion of Science for Young Scientists (No. 26$\cdot$2428).


\end{document}